\documentclass[aps,prl,twocolumn,groupedaddress,nofootinbib]{revtex4}

\usepackage[dvips]{graphicx}
\usepackage{color}
\usepackage{amsmath,amssymb,slashed}
\usepackage{tikz}

\newif\ifdraft
\drafttrue
\newif\ifpreprint
\preprinttrue

\def\spa#1.#2{\left\langle#1\,#2\right\rangle}
\def\spb#1.#2{\left[#1\,#2\right]}

\font\tenshuffle=shuffle10 \font\sevenshuffle=shuffle7 \font\fiveshuffle=shuffle7 at 5pt
\def\shuffle{{%
\def\Dshuffle{\mathbin{\hbox{\tenshuffle\char'001}}}%
\def\Sshuffle{\mathbin{\hbox{\sevenshuffle\char'001}}}%
\def\SSshuffle{\mathbin{\hbox{\fiveshuffle\char'001}}}%
\mathchoice{\Dshuffle}{\Dshuffle}{\Sshuffle}{\SSshuffle}}}

\newcommand{\eq}{\begin{equation}}
\newcommand{\eqe}{\end{equation}}
\newcommand{\eqa}{\begin{eqnarray}}
\newcommand{\eqae}{\end{eqnarray}}

\newcommand{\bea}{\begin{eqnarray}}
\newcommand{\eea}{\end{eqnarray}}
\newcommand{\dd}{\mathrm{d}}

\newbox\charbox
\newbox\slabox
\def\s#1{{      
        \setbox\charbox=\hbox{$#1$}
        \setbox\slabox=\hbox{$/$}
        \dimen\charbox=\ht\slabox
        \advance\dimen\charbox by -\dp\slabox
        \advance\dimen\charbox by -\ht\charbox
        \advance\dimen\charbox by \dp\charbox
        \divide\dimen\charbox by 2
        \raise-\dimen\charbox\hbox to \wd\charbox{\hss/\hss}
        \llap{$#1$}
}}

\begin{document}

\title{
New Relations for Gauge-Theory and Gravity Amplitudes at Loop Level
} 

\author{Song He$^a$ and
Oliver Schlotterer$^{b}$}
\affiliation{$^a$ CAS Key Laboratory of Theoretical Physics, Institute of Theoretical Physics \& UCAS, 
Chinese Academy of Sciences, Beijing 100190, China
}
\affiliation{$^b$ Max-Planck-Institut f\"ur Gravitationsphysik, Albert-Einstein-Institut,
14476 Potsdam, Germany.}

\begin{abstract}
In this letter, we extend the tree-level Kawai--Lewellen--Tye (KLT) and Bern--Carrasco--Johansson (BCJ) amplitude relations to loop integrands of gauge theory and gravity. 
By rearranging the propagators of gauge and gravity loop integrands, we propose the first manifestly gauge- and diffeomorphism invariant formulation of their double-copy relations. 
The one-loop KLT formula expresses gravity integrands in terms of more basic gauge invariant building blocks for gauge-theory amplitudes, dubbed {\it partial integrands}. 
The latter obey a one-loop analogue of the BCJ relations, and both KLT and BCJ relations are universal to bosons and fermions in any number of spacetime dimensions and independent on the amount of supersymmetry.
Also, one-loop integrands of Einstein--Yang--Mills (EYM) theory are related to partial integrands of pure gauge theories.
%
\end{abstract}

\maketitle

\section{Introduction}  

\vspace{-0.3cm}
\noindent
A unified perspective on fundamental forces intertwining gravity with gauge interactions is suggested by string theory: Gravitons arise as the massless vibration modes of closed strings which are in turn formed by joining the endpoints of open strings with gauge bosons among their ground-state excitations. A prominent perturbative manifestation of gravity's resulting double-copy structure was revealed by Kawai, Lewellen and Tye (KLT) in 1985: The {\it KLT formula} \cite{Kawai:1985xq} assembles the tree-level S-matrix of closed strings from squares of color-stripped open-string amplitudes. Accordingly, its point-particle limit relates tree amplitudes of Einstein gravity (and its supersymmetric extensions) to squares of gauge-theory partial amplitudes. The structure of the KLT formula turned out to apply universally to tree amplitudes in a variety of non-gravitational theories, opening up a double-copy perspective on Born--Infeld theory and special Galileons \cite{Cachazo:2014xea}, as well as, surprisingly, even the open string \cite{Mafra:2011nv, Broedel:2013tta, Carrasco:2016ldy}.

It is remarkable that the double-copy structure appears to extend to the quantum regime \cite{Bern:2010ue}, as supported by impressive constructions of multiloop supergravity amplitudes from gauge-theory input such as \cite{Bern:2012uf, Bern:2013uka}. The unprecedented efficiency of this method gives rise to hope that it allows to pinpoint the onset of ultraviolet divergences in various supergravity theories, see e.g.~\cite{Bern:2013uka}. So far, such double-copy constructions have been carried out at the level of cubic diagrams which have to be represented in a particular gauge of the spin-one constituents and thereby obscure the diffeomorphism invariance of gravity amplitudes. In this letter, we close this gap and give the first manifestly gauge- and diffeomorphism invariant ``KLT-like'' double-copy formula for their loop integrands. It does not depend on any particular gauge in arranging the cubic-diagram representation of the gauge-theory integrands, and it takes a universal form for bosons and fermions, regardless of the number of spacetime dimensions and supersymmetries. 

The double-copy approach to perturbative gravity relies on a hidden symmetry of the gauge-theory S-matrix -- the duality
between color and kinematics due to Bern, Carrasco and Johansson (BCJ) \cite{Bern:2008qj}. Similar to the KLT formula, the manifestly gauge invariant tree-level incarnation of this duality known as the {\it BCJ relations} among color-stripped amplitudes was initially derived from string theory \cite{BjerrumBohr:2009rd} and turns out to also apply to effective scalar theories including non-linear sigma models (NLSM) \cite{Chen:2013fya, Carrasco:2016ldy}. Generalizations of BCJ relations to one-loop integrands have already been given in a field-theory \cite{Boels:2011tp} and string-theory \cite{Tourkine:2016bak} context, and we will provide an alternative formulation which is tailored to play out with a KLT formula for one-loop gravity integrands.

A convenient framework to complement the string-theory perspective on double copies as well as KLT and BCJ relations is the {\it CHY formalism} due to Cachazo, Yuan and one of the current authors \cite{Cachazo:2013iea}. The CHY prescription for loop amplitudes \cite{Adamo:2013tsa,Geyer:2015bja} manifests their relation with forward limits of tree-level building blocks \cite{Cachazo:2015aol} where one sums over the polarization and color degrees of freedom of two extra legs with back-to-back momenta~\cite{forwardlimit}. Indeed, the implementation of forward limits in \cite{Geyer:2015bja,Cachazo:2015aol} led us to identify the main results of this letter, as will be detailed in \cite{toappear}. By the same argument, we find similar results for Einstein--Yang--Mills (EYM) theory. Given the wide range of double-copy theories, our results 
should extend to the NLSM and its double copies and
reveal universal structural insights into the quantum regime of perturbative field and string theory.
\vspace{-.5em}
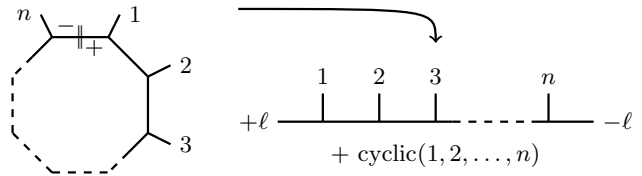
\begin{figure}[h]
\begin{center}
\begin{tikzpicture} [scale=0.75, line width=0.30mm]
\scope[xshift=-0.8cm]
\draw (0.5,0)--(-0.5,0);
\draw (-0.5,0)--(-0.85,-0.35);
\draw [dashed](-0.85,-0.35)--(-1.2,-0.7);
\draw (0.5,0)--(1.2,-0.7);
\draw[dashed] (-1.2,-1.7)--(-1.2,-0.7);
\draw (1.2,-1.7)--(1.2,-0.7);
\draw (1.2,-1.7)--(0.85,-2.05);
\draw[dashed] (0.85,-2.05)--(0.5,-2.4);
\draw[dashed] (-0.5,-2.4)--(0.5,-2.4);
\draw[dashed] (-0.5,-2.4)--(-1.2,-1.7);
\draw (-0.5,0)--(-0.7,0.4)node[left]{$n$};
\draw (0.5,0)--(0.7,0.4)node[right]{$1$};
\draw (1.2,-0.7)--(1.6,-0.5)node[right]{$2$};
\draw (1.2,-1.7)--(1.6,-1.9)node[right]{$3$};
\draw (0,0) node{$| \! |$};
\draw (-0.25,0.2)node{$-$};
\draw (0.25,-0.2)node{$+$};
\endscope
\draw[->] (2,0.5) .. controls (5.5,0.5) .. (5.5,-0.1);
\draw(5.5,-2.1)node{$+ \ {\rm cyclic}(1,2,\ldots,n)$};
\scope[yshift=0.5cm]
\draw (2.7,-2)node[left]{$+\ell$} -- (5.8,-2);
\draw (7.2,-2) -- (8.3,-2)node[right]{$-\ell$};
\draw (3.5,-2) -- (3.5,-1.5)node[above]{$1$};
\draw (4.5,-2) -- (4.5,-1.5)node[above]{$2$};
\draw (5.5,-2) -- (5.5,-1.5)node[above]{$3$};
\draw[dashed] (5.8,-2) -- (7.2,-2);
\draw (7.5,-2) -- (7.5,-1.5)node[above]{$n$};
%
\endscope
\end{tikzpicture}
\caption{Decomposing $n$-gon integrals into $n$ tree diagrams.}
\label{fig1}
\end{center}
\end{figure}

\section{Main results}

\vspace{-0.3cm}

{\bf A. Definition of loop integrands:}  The key ingredient in our construction is a new representation of loop integrands in gauge and gravity theories, first considered in~\cite{Geyer:2015bja} and~\cite{Baadsgaard:2015twa}. The new representation can be obtained by first rearranging any Feynman loop integrand via partial-fraction relations and then shifting the loop momentum, which will not change the integrated result in dimensional regularization. At one-loop level, this procedure converts an $n$-gon integral into a sum of $n$ terms, where all the inverse ``propagators" but one become linear in the loop momentum. For example, the $n$-gon scalar integral in figure \ref{fig1} can be written as
\begin{align}
\int& \frac{2^{n-1} \  \dd^D L }{L^2 (L{+}k_1)^2 \! \ldots \! (L{+} k_{1}{+}k_{2}{+}\ldots {+}k_{n-1})^2} =\int \frac{\dd^D \ell}{\ell^2}  \notag\\
& \! \! \! \times  \bigg\{ \frac{ 1}{s_{1,\ell} s_{12 ,\ell} \ldots s_{12\ldots n-1 ,\ell}} 
+{\rm cyclic}(1,2,\ldots,n) \bigg\} \label{ngon} 
\end{align}
with inverse propagators $s_{i j \ldots}\equiv \frac 1 2 (k_i{+}k_j{+}\ldots)^2$ and 
\begin{equation}
s_{i j \ldots, \ell}\equiv \ell \cdot(k_i{+}k_j{+}\ldots)+ \tfrac{1}{2} (k_i{+}k_j{+}\ldots)^2 \ .
\end{equation}
The procedure can be applied to any one-loop amplitude with local propagators, and the result can be identified as tree diagrams involving two off-shell legs with momenta $\pm \ell$~\cite{Cachazo:2015aol}. For non-supersymmetric theories, tree amplitudes diverge in the forward limit, but the divergences can be regulated using the prescription of~\cite{Baadsgaard:2015twa}, or that of~\cite{Cachazo:2015aol} in the CHY representation. 

In this way, one can rewrite one-loop $n$-point gravity amplitudes $M_n$ in the new representation as
\begin{align}
M_n&
=\int \frac{\dd^D \ell}{\ell^2}~m_n (\ell)\,.
\label{gravint}
\end{align} 
We will refer to $m_n (\ell)$ as the {\it integrand} for one-loop gravity amplitudes -- its $n{-}1$ propagators are linear in $\ell$ after stripping off the overall $1/\ell^2$, and it can be obtained from a KLT formula to be spelt out below.

\medskip
{\bf B. Partial integrands:} The backbone of our KLT formula for $m_n (\ell)$ is a novel refinement of color-stripped gauge-theory amplitudes $A(1,2,\ldots,n)$. Given the rearrangement of loop integrals in (\ref{ngon}), it is natural to collect all terms with the loop momentum flowing from leg $i$ to $i{+}1$ (cf.\ figure \ref{fig1}) in the more basic building block $a(1,2,\ldots,i,-,+, i{+}1, \ldots, n)$. Identifying $k_{\pm} \equiv \pm \ell$, this {\it partial integrand} can be viewed as a tree amplitude with cyclic ordering $(1,2,\ldots,i,-,+, i{+}1, \ldots, n)$ in the forward limit of two off-shell legs $-$ and $+$. Hence, partial integrands are individually gauge invariant, assuming that forward-limit divergences are suitably regulated. By definition, the sum of $n$ such partial integrands gives the complete integrand of one-loop color-ordered amplitudes:  
\begin{align}\label{singtrac}
A(1,2,\ldots,n)=\int \frac{\dd^D \ell}{\ell^2}
 \sum_{i=1}^n  a(1,2,\ldots,i,-,+, i{+}1, \ldots, n)\,. 
\end{align}
The notion of partial integrands is naturally motivated by the one-loop formula in \cite{Geyer:2015bja}, but we emphasize that the definition here is general and independent of any particular representation. Moreover, the decomposition \eqref{singtrac} of color-ordered amplitudes differs significantly from Q-cut approach in eq.\ (6) of \cite{Baadsgaard:2015twa}. In the following we propose BCJ relations among these $a$'s, and KLT relations that combine them to the gravity integrands $m_n$ in \eqref{gravint}. 

\medskip
{\bf C. One-loop BCJ relations:} Having defined partial integrands of the form $a(\pi (1,2,\ldots,n), -,+)$ with permutations $\pi \in S_{n}$, one defines $a(\alpha, -, \beta, +)$ for non-adjacent legs $-$ and $+$ (with multiparticle labels such as $\alpha=\{\alpha_1, \alpha_2,\ldots,\alpha_p\}$) via Kleiss--Kuijf relations of the underlying $(n{+}2)$-point trees \cite{Kleiss}. We conjecture that they further satisfy universal BCJ relations like $(n{+}2)$-point trees \cite{Bern:2008qj}, leaving at most $(n{-}1)!$ independent partial integrands, even though additional relations may exist for special theories. These relations can be generated by fundamental BCJ relations which involve changing the position of one leg only. For instance, choosing $+$ or $1$ and defining $k_{12\ldots i} \equiv k_1+k_2+\ldots+k_i$ yields
\begin{align}
\sum_{i=1}^{n{-}1} (\ell \cdot k_{12\ldots i}) ~a(1,2,\ldots, i, +, i{+}1,\ldots, n,-)=0\,,
\label{BCJ1}
\end{align} 
and, by a similar use of momentum conservation,
\begin{align}
&\sum_{i=2}^{n{-}1} (k_1 \cdot k_{23\ldots i})
~a(2,3,\ldots, i, 1, i{+}1, \ldots, n, -, +) \notag \\
& \ \ \ = (\ell\cdot k_1) ~a(2,3,\ldots,n,-,1,+)\,.
\label{BCJ2}
\end{align}
The BCJ relations should hold universally in $D$ spacetime dimensions, for external bosons and fermions in the adjoint representation of the gauge group and independently on the extent of supersymmetry in $a, \tilde a$.

\medskip
{\bf D. One-loop KLT relations:} As our main result, we conjecture that the gravity integrands $m_n(\ell)$ in (\ref{gravint}) can be obtained as bi-linears of partial integrands $a,\tilde a$ of (possibly different) gauge theories. This manifestly gauge- and diffeomorphism invariant incarnation of the double copy \cite{Bern:2010ue} matches the $(n{+}2)$-point tree-level KLT relations \cite{Kawai:1985xq} with two additional legs $k_{\pm}{=}\pm \ell$,
\begin{align}
m_n=\sum_{\pi, \rho\in S_{n{-}1}} a(+, \pi, n, -)~S[\pi|\rho]_{\ell}~\tilde a(+, \rho, -, n)\,,
\label{1KLT}
\end{align}
where $\pi, \rho$ are permutations of $1,2,\ldots, n{-}1$. The all-multiplicity KLT matrix $S[\pi|\rho]_{\ell}$ identified in \cite{Bern:1998sv} has been studied in the momentum-kernel formalism \cite{BjerrumBohr:2010hn} and allows for the recursive definition \cite{Carrasco:2016ldy}
\begin{equation}
S[\alpha,j | \beta,j,\gamma]_\ell = k_j  \cdot (\ell{+}k_{\beta})  \, S[\alpha | \beta,\gamma]_\ell
\, , \ \ \  S[\emptyset | \emptyset]_\ell = 1 \, ,
\end{equation}
with $k_\beta=k_{\beta_1}+\ldots+k_{\beta_p}$ and $\beta,\gamma=\{\beta_1,\ldots,\beta_p,\gamma_1,\ldots \gamma_q\}$ denoting the composition of multiparticle labels $\beta$ and~$\gamma$. The KLT relations (\ref{1KLT}) should be valid for any (possibly asymmetric) double copy of gauge theories whose partial integrands satisfy the above one-loop BCJ relations.

\section{Examples}

\vspace{-0.3cm}
\noindent
In this section, we provide evidence for one-loop BCJ and KLT relations, using examples of (partial) integrands $a$ and $m_n$ in gauge theories and supergravities with maximal or half-maximal supersymmetry. Supergravity integrands with 24 supercharges can be obtained from the double copy of $a^{1/2}$ and $\tilde a^{\rm max}$. We exemplify the degenerations of the BCJ basis in these supersymmetric cases below $(n{-}1)!$ elements and leave explicit checks for theories without supersymmetry to the future.

\medskip
{\bf A. 4pt maximal:} One-loop four-point amplitudes of $D$-dimensional SYM with 16 supercharges are
determined by the permutation invariant $t_8$-tensor \cite{GSB}
\begin{align}
&t_8(A,B,C,D) \equiv f^{mn}_Af^{np}_B f^{pq}_C f^{qm}_D  \\
&- \frac{1}{4} f^{mn}_Af^{nm}_B f^{pq}_C f^{qp}_D + {\rm cyc}(B,C,D) \notag
\end{align}
with polarization vectors $e_i^m$, linearized field strength $f_i^{mn}\equiv k^m_i e^n_i - k^n_i e^m_i$ and $D$-dimensional vector indices $m,n=0,1,\ldots,D{-}1$. The partial integrands include
\begin{align}
a^{\rm max}(1,2,3,4,-,+) &= \frac{t_8(1,2,3,4)}{s_{1,\ell} s_{12,\ell} s_{123,\ell}} \label{max4} \\
a^{\rm max}(1,2,3,-,4,+) &=  t_8(1,2,3,4) \bigg[ \frac{1}{s_{1,\ell} s_{12,\ell} s_{4,\ell}}
 \\
&\! \! \! \! \! \! \! \! \! \! \! \! \! \! \! \! \! \! \! \! \! \! \! \! \! \! \! \! \! \! \! \! \! \! \! \! 
+\frac{1}{s_{1,\ell} s_{12,\ell} s_{3,\ell}} 
+\frac{1}{s_{1,\ell} s_{14,\ell} s_{3,\ell}}
+\frac{1}{s_{4,\ell} s_{14,\ell} s_{3,\ell}} \bigg] \notag
\label{teight}
\end{align}
and a similar six-term formula for $a^{\rm max}(1,2,-,3,4,+)$. One can check that the supergravity integrand
\begin{align}
m^{\rm max}_4(\ell)=   \frac{ |t_8(1,2,3,4)|^2}{s_{1,\ell} s_{12,\ell} s_{123,\ell}} 
+{\rm perm}(1,2,3,4)
\end{align}
adapted to 32 supercharges follows from the KLT formula
\begin{align}
&m^{\rm max}_4(\ell)= - \sum_{\pi,\rho \in S_3} a^{\rm max}(\pi(1,2,3),4,-,+)  \\
& \ \times S[\pi(1,2,3)|\rho(1,2,3)]_{\ell}  \ \tilde a^{\rm max}(\rho(1,2,3),-,4,+)  \ .\notag
\end{align}
Note that the universal kinematic factor $t_8(1,2,3,4)$ selected by maximal supersymmetry only leaves one linearly independent partial integrand, going {beyond} the naively six-dimensional BCJ basis of $a(\pi(1,2,3),4,-,+)$.

\medskip
{\bf B. 3pt half-maximal:} The infrared regularization prescription of \cite{BBS} for SYM amplitudes in $D\leq 6$ with 8 supercharges gives rise to three-point partial integrands
\begin{align}
&a^{1/2}(1,2,3,-,+) = \frac{ \ell_m \big[e^m_1(k_2\cdot e_3)(k_3\cdot e_2) {+} (1{\leftrightarrow} 2,3) \big] }{s_{1,\ell} s_{12,\ell}}  \notag \\
 & \ \ \ \ \ \ \ \ \ \  \ \ \  - \frac{(e_1\cdot e_2)(k_1\cdot e_3)}{s_{12,\ell}}
 - \frac{(e_2\cdot e_3)(k_2\cdot e_1)}{s_{1,\ell}} \ ,
\end{align}
where the bubble numerators $s_{ij}(e_i\cdot e_j) (k_i\cdot e_p)$ compensate for the divergent propagators $s_{ij}^{-1}$, and the triangle numerator vanishes upon $\ell^m \rightarrow k_j^m$. Kleiss--Kuijf relations \cite{Kleiss} and momentum conservation yield
\begin{align}
&a^{1/2}(+,1,2,-,3) = 0 \ ,
\label{zero}
\end{align}
such that the KLT formula with one factor of $\tilde a^{1/2}(\rho(1,2),-,3,+)$  in each term identifies a vanishing supergravity integrand, $m^{1/2}_3(\ell) = 0 $.

\medskip
{\bf C. 5pt maximal:} The $(D\leq 10)$-dimensional bosonic components of the five-point results in pure-spinor superspace \cite{Mafra:2014gja} yield local and gauge invariant expressions~\cite{private}
\begin{align}
&a^{\rm max}(1,2,3,4,5,-,+) = \frac{N_{12345}^{\ell}}{s_{1,\ell} s_{12,\ell} s_{123,\ell} s_{1234,\ell}} \notag \\
&- \frac{t_8(12,3,4,5) }{s_{12} s_{12,\ell} s_{123,\ell} s_{1234,\ell} } 
- \frac{ t_8(1,23,4,5)} {s_{23} s_{1,\ell} s_{123,\ell}  s_{1234,\ell}} \label{cyc12345}  \\
&- \frac{ t_8(1,2,34,5)}{ s_{34} s_{1,\ell}  s_{12,\ell} s_{1234,\ell}}
- \frac{t_8(1,2,3,45)} {s_{45} s_{1,\ell} s_{12,\ell} s_{123,\ell}}  \notag
\end{align}
with pentagon numerator 
\begin{align}
&N_{12345}^{\ell } = \ell_m\big[ e_1^m t_8(2,3,4,5) + (1 \leftrightarrow 2,3,4,5) \big] \! \! \! \! \!
\\
&- \tfrac{1}{2}\big[ t_8(12,3,4,5) {+} (12{\leftrightarrow} 13,14,15,23,24,25,34,35,45) \big] \notag
\end{align}
and two-particle field-strength $f^{mn}_{12}$ in $t_8(12,3,4,5)$,
\begin{align}
e^{m}_{12}  &\equiv  e_2^m (k_2\cdot e_1) -  e_1^m (k_1\cdot e_2) +\tfrac{1}{2}(k_1^m - k_2^m) (e_1 \cdot e_2)
\notag\\
f^{mn}_{12}&\equiv k_{12}^m e_{12}^n - k_{12}^n e_{12}^m
-s_{12} ( e_1^m e_2^n - e_1^n e_2^m)   \ .
\label{BG32}
\end{align}
In the full single-trace integrand (\ref{singtrac}), the propagators $(\ell{-}k_{12\ldots j})^2$ quadratic in $\ell$ can be recovered from (\ref{cyc12345}) by the following property of the vector pentagon:
\begin{align}
& N_{12345}^{\ell \rightarrow 0} -N_{12345}^{\ell \rightarrow k_1} = t_8(12,3,4,5) + (2\leftrightarrow 3,4,5) 
\end{align}
The supergravity integrand $m_5^{\rm max}$ following from the KLT relation (\ref{1KLT}) is checked to reproduce the $5!$ pentagon diagrams and $10\cdot 4!$ box diagrams expected from the BCJ duality and double-copy \cite{Bern:2010ue, Carrasco:2011mn, Mafra:2014gja}. Note that the naively 24-dimensional BCJ basis of partial integrands is again degenerate by maximal supersymmetry -- the gauge invariant organization of \cite{Mafra:2014gja} leaves three linearly independent kinematic factors, two permutations of $A^{\rm tree}_{\rm YM}(1,2,3,4,5)$ and one $\ell$-dependent invariant.

\medskip
{\bf D. 4pt half-maximal:} In the four-point SYM integrands of \cite{BBS}, all the non-trivial tree-level diagrams involving leg 1 are absorbed into the gauge invariant quantities $C_{1|ijk},C^m_{1|ij,k},C^{mn}_{1|i,j,k}$ defined in the reference,
\begin{align}
&a^{1/2}(1,2,3,4,-,+) = \frac{  C_{1|234} }{s_{1,\ell}}
-\frac{ \ell_m C^m_{1|23,4} }{s_{1,\ell} s_{123,\ell}}
-\frac{ \ell_m C^m_{1|34,2} }{s_{1,\ell} s_{12,\ell}} \notag \\
&\!  + \frac{\ell_m \ell_n C^{mn}_{1|2,3,4} - \ell_m\big[ s_{23} C^m_{1|23,4} + (23 \leftrightarrow  24,34)\big] }{2 s_{1,\ell} s_{12,\ell} s_{123,\ell} } \ . \! \! \! \!
\label{4halfmax}
\end{align}
This expression is valid in $D\leq 6$ and arises from solely the hypermultiplet running in the loop. 
Contributions from additional vector multiplets can be obtained by linear combinations of (\ref{4halfmax}) and its maximally supersymmetric counterpart (\ref{max4}). 
BCJ relations and reflection properties can be checked through the permutations
\begin{align}
C^m_{2|13,4} &= C^m_{1|32,4} + k_4^m C_{1|324} \\
C^m_{2|34,1} &= C^m_{1|34,2}+\big[ C^m_{1|23,4}+ k_4^m C_{1|234} - (3\leftrightarrow 4)\big] \notag \\
C^{mn}_{2|1,3,4} &= C^{mn}_{1|2,3,4}
{+}
2\big[ k_{3}^{(m}C^{n)}_{1|23,4}{+}k_{3}^{(m} k_4^{n)} C_{1|234} {+} (3{\leftrightarrow }4)\big]  \notag
 \\
& \ \ \ \ \ \ \ \ +  2i \eta^{mn} \epsilon_{pqrstu} e^{p}_{1} e^q_{2} k_3^{r}e_3^{s} k_4^{t} e^{u}_4 \ , \notag
\end{align}
up to the fingerprints of the $D=6$ box anomaly in the last line.
The $\ell$-dependent invariants such as $\ell_m C^m_{1|23,4}$ leave several linearly independent partial integrands, and the role of the anomaly from $\ell_m \ell_n C^{mn}_{1|2,3,4}$ in the KLT formula for $m^{1/2}_4$ will be investigated in the future \cite{toappear}.

\medskip
{\bf E. MHV maximal:} BCJ numerators for one-loop MHV amplitudes in ${\cal N}=4$ SYM have been given at all multiplicities \cite{He:2015wgf} in terms of $X_{i,j} \equiv \langle 1 |  \! \not \! k_i  \! \not \! k_j \, | 1 \rangle$~\cite{Monteiro:2011pc}. The four- and five-point partial integrands read
\begin{align}
a^{\rm MHV}(1,2,3,4,-,+) &= \frac{ \delta^8(Q) }{\prod_{j=2}^4\langle 1j\rangle^2} \frac{ X_{2,3}^2 }{s_{1,\ell} s_{12,\ell} s_{123,\ell} } \\
a^{\rm MHV}(1,2,{\ldots},5,-,+) &= \frac{ \delta^8(Q) }{\prod_{j=2}^5\langle 1j\rangle^2} \Big\{ 
\frac{ X_{3,2} X_{4,5}^2   }{s_{23} s_{1,\ell} s_{123,\ell} s_{1234,\ell} }  \notag \\
& \! \! \! \! \! \! \!  \! \! \! \! \! \! \! \! \! \! \! \! \! \!  \! \! \! \! \! \! \!  \! \! \! \! \! \! \! \! \! \! \! \! \! \!  \! \! \! \! \! \! \! + \frac{ X_{4,3}  X_{2,5}^2  }{s_{34} s_{1,\ell} s_{12,\ell} s_{1234,\ell} }   + \frac{ X_{5,4} X_{2,3}^2  }{s_{45} s_{1,\ell} s_{12,\ell} s_{123,\ell} }   \\
& \! \! \! \! \! \! \!  \! \! \! \! \! \! \!  \! \! \! \! \! \! \!  \! \! \! \! \! \! \!  \! \! \! \! \! \! \! \! \! \! \! \! \! \!  \! \! \! \! \! \! \! +
\frac{ X_{2,4}X_{2,3}X_{\ell,5}+X_{2,5}X_{2,3}X_{4,5}+X_{3,5}X_{\ell,2}X_{4,5} }{  s_{1,\ell} s_{12,\ell} s_{123,\ell}  s_{1234,\ell} } 
\Big\} \ , \notag
\end{align}
with the standard super-momentum conserving delta function $\delta^{8}(Q)$ \cite{Nair:1988bq},
and the $n$-point generalization can be straightforwardly extracted from \cite{He:2015wgf}. We have checked up to $n{=}10$ explicitly, and expect to have a all-multiplicity proof, that our one-loop KLT formula \eqref{1KLT} reproduces the MHV supergravity integrands of~\cite{He:2015wgf}.

\vspace{-0.4cm}

\section{Implications for Einstein--Yang--Mills}

\vspace{-0.3cm}

\noindent
Given the one-loop KLT relations (\ref{1KLT}) between amplitudes of pure (super-)gravity and pure gauge theories, it is natural to investigate their minimal coupling within EYM theories. At tree level, EYM amplitudes were related to pure gauge-theory amplitudes \cite{eymref1,eymref2}, and we will give one-loop extensions of such relations in this section.


\medskip
{\bf A. Partial integrands for EYM:} In analogy to (\ref{singtrac}), one can decompose the single-trace sector of one-loop EYM amplitudes with insertions of $m$ external gravitons $\{p\} = \{p_1,p_2,\ldots,p_m\}$ into partial integrands $a_{\rm EYM}$,
\begin{align}
&A_{\rm EYM}(1,2,\ldots,n; \{p\})=\int \frac{\dd^D \ell}{\ell^2}
\label{EYM1} \\
& \,\times  \sum_{i=1}^n a_{\rm EYM}(1,2,\ldots,i,-,+, i{+}1, \ldots, n; \{p\}) \ ,
\notag
\end{align}
where any propagator in the second line is rendered linear in $\ell$ via (\ref{ngon}). With the convention that (\ref{EYM1}) only tracks the propagation of gauge multiplets in the loop, any $a_{\rm EYM}$ can be obtained from the forward limit of a tree-level EYM amplitude with single-trace ordering $(1,2,\ldots,i,-,+, i{+}1, \ldots, n)$. Extensions of (\ref{EYM1}) to incorporate graviton propagators are related to trees with additional $p_j$ and go beyond the scope of this work.

\medskip
{\bf B. One-loop amplitude relations:} The forward limit of the tree-level amplitude relations such as \cite{eymref1}
\begin{align}
&A_{\rm EYM}^{\rm tree}(1,2,\ldots,n; p) = \sum_{j=1}^{n-1} (e\cdot k_{12\ldots j})  \notag \\
&\ \ \ \ \ \times A^{\rm tree}_{\rm YM}(1,2,\ldots,j,p,j{+}1,\ldots,n)
\label{EYM2} 
\end{align}
for a single graviton implies the one-loop identity
\begin{align}
&a_{\rm EYM}(+,1,2,\ldots,n,-; p) = -(e\cdot \ell) \,
a(+,1,2,\ldots,n,-,p)
\notag \\
&\ \ +\sum_{j=1}^{n-1} (e\cdot k_{12\ldots j}) \,  a(+,1,2,\ldots,j,p,j{+}1,\ldots,n,-)
\label{EYM3} 
\end{align}
among partial integrands
whose gauge invariance follows from the BCJ relation (\ref{BCJ2}). The maximally supersymmetric four-point instance can be easily checked to descend from three box integrals:
\begin{align}
&A_{\rm EYM}^{\rm max}(1,2,3; p) =  t_8(1,2,3,p)  \label{EYM5} \\
& \times \int \frac{8 \,  \dd^D L \  \, (e \cdot L) }{L^2 (L{+}k_{123})^2 (L{+}k_{23})^2 (L{+}k_{3})^2}  + {\rm cyc}(1,2,3) \, . \notag
\end{align}
A similar identification of legs $(1,n) \rightarrow (+,-)$ can be performed to promote the results of \cite{eymref2} with additional graviton insertions to loop level, resulting for instance in the two-graviton example (\ref{EYM4}) in the appendix. Such amplitude relations take a universal form, irrespective of the supersymmetries preserved by $a$ or $a_{\rm EYM}$, and a CHY derivation will be given in \cite{toappeartwo}. It would be interesting to convert the multitrace results of \cite{eymref2} into one-loop contributions to EYM amplitudes with graviton propagators.

\vspace{-0.4cm}

\section{Conclusions and further directions}

\vspace{-0.3cm}
\noindent
In this letter, we have identified partial integrands as basic gauge invariant building blocks for one-loop gauge-theory amplitudes; they arise naturally from the new representation of one-loop amplitudes \cite{Geyer:2015bja, Baadsgaard:2015twa}, such as (\ref{ngon}) of $n$-gon integrals, and can be derived using forward limits of tree amplitudes, or CHY representations~\cite{Cachazo:2015aol}. These partial integrands inherit BCJ relations (\ref{BCJ1}), (\ref{BCJ2}) from tree level, and similar relations for partial integrands of one-loop EYM amplitudes (\ref{EYM2}). Most importantly, they are suitable for constructing one-loop gravity integrands through our main result (\ref{1KLT}), a one-loop generalization of the KLT formula \cite{Kawai:1985xq} valid for the gravitational double copies of a wide range of gauge theories. 

Furthermore, the notion of partial integrands and their one-loop BCJ relations directly carry over to NLSM amplitudes. Parallel to tree-level case~\cite{Cachazo:2014xea,Cachazo:2016njl}, the one-loop KLT formula with two copies of NLSM gives the one-loop integrand of special Galileons, and that with NLSM and (super-)Yang--Mills theory yields integrands of Born--Infeld theory along with supersymmetric extensions to Dirac--Born--Infeld--Volkov--Akulov theories~\cite{Volkov:1972jx}.

The one-loop KLT formula is manifestly gauge- and diffeomorphism invariant, and it holds regardless of any particular representation of gauge-theory partial integrands. Once a gauge-theory amplitude has been expressed in accordance with the duality between color and kinematics \cite{Bern:2008qj}, one can view the formula (\ref{1KLT}) as reorganizing the cubic diagrams in the double-copy representations of \cite{Bern:2010ue} into gauge invariant building blocks. In absence of duality-satisfying representations, however, (\ref{1KLT}) yields supergravity integrands which have been previously out of reach. In order to take maximal advantage of the KLT formula, it remains to systematically develop integration routines for rearranged loop integrals in $m_n$ (see~\cite{Baadsgaard:2015twa} for an example of four-point one-loop integration relevant to both partial integrands and Q-cuts). It should be feasible to reinstate the standard propagators quadratic in the loop momenta through an algorithmic procedure. On the other hand, it is highly desirable to directly extract physical information, such as unitarity cuts or ultraviolet divergences of one-loop amplitudes, from the new representation of the integrands. 

Although the discussion has been adapted to the one-loop case, we expect that the notion of partial integrands and their KLT composition to permutation invariant gravity integrands extends to any loop order. For instance, we have identified all two-loop four-point partial integrands in maximal SYM, obtained from double forward limits of eight-point tree amplitudes \cite{Geyer:2016wjx}. These naturally lead to a new proposal for maximal supergravity integrands through the corresponding KLT formula, which is similar to that of eight-point trees. We leave it to future work to verify the proposal and to gather more evidence for higher loops and multiplicities.

\vspace{-0.4cm}

\section{Acknowledgements}

\vspace{-0.3cm}

We would like to thank Nima Arkani-Hamed, Marcus Berg, Igor Buchberger, Henrik Johansson, Carlos Mafra, Dhritiman Nandan, Jan Plefka, Lorenzo Tancredi, Piotr Tourkine, Congkao Wen, Ellis Ye Yuan and in particular Johannes Br\"odel for combinations of inspiring discussions, valuable comments on a draft of the manuscript and collaboration on related topics. O.~S.~is grateful to the Institute of Theoretical Physics, CAS in Beijing for hospitality during initial stages of this work and to Caltech and UCLA for hospitality during its finalization. S.~H.~thanks the Institute for Advanced Study for hospitality during the finalization. S.H.’s research is supported in part by the
Thousand Young Talents program and the Key Research Program of Frontier Sciences of
CAS.

\vspace{-0.4cm}

\section{Appendix: EYM with two gravitons}

\vspace{-0.3cm}
\noindent
Starting from the relations for single-trace EYM tree amplitudes with two gravitons \cite{eymref2}, we relabel two gluon legs $(1,n) \rightarrow (+,-)$ with momenta $k_{\pm}=\pm\ell$ to obtain
\small
\begin{align}
&a_{\rm EYM}(+,1,2,\ldots,n,-; p,q) = \sum_{1=i\leq j}^{n-1} (e_p \cdot k_{12\ldots i})(e_q \cdot k_{12\ldots j}) 
\notag \\
& \ \ \ \ \times a(+,1,2,\ldots,i,p,i{+}1,\ldots,j,q,j{+}1,\ldots, n ,-) \notag 
\\
&+ (e_p\cdot \ell)  (e_q \cdot \ell) a(-,p,q,+,1,2,\ldots,n) \notag \\
&+(e_p\cdot \ell) (e_q \cdot p) \big[  a(-,q,p,+,1,\ldots,n) + a(q,-,p,+,1,\ldots,n)  \big] \notag \\
&-(e_p \cdot \ell) \sum_{j=1}^{n-1} (e_q \cdot k_{12\ldots j}) a(-,p,+,1,\ldots,j,q,j{+}1,\ldots,n)
\notag  \\
&-(e_q\cdot p) \sum_{j=1}^{n-1} (e_p \cdot k_{12\ldots j}) \Big[ a(+,1,2,\ldots,j,p,j{+}1,\ldots,n,q,-)
 \notag \\
&\ \ \ \ \ \ \ \ \ \ \ \ +a( \{q \shuffle +,1,2,\ldots,j\},p,j{+}1,\ldots,n,-)  \Big] \notag \\
&-\frac{1}{2}(e_p\cdot e_q) \Big[ \sum_{i=1}^{n}(p\cdot k_i) a( \{q,p \shuffle +,1,\ldots,i{-}1\},i ,i{+}1,\ldots,n,-) \notag \\
&\ \ \ \ \ \ \ \ \ \ \ \ + (p\cdot \ell) a(q,p,+,1,\ldots,n,-) \Big] + (p\leftrightarrow q) \ ,
\label{EYM4} 
\end{align} \normalsize
where $\{ \beta \shuffle \gamma\}$ denotes the shuffle product. This reduces 
partial integrands (\ref{EYM1}) for one-loop EYM amplitudes with two graviton insertions 
to partial integrands (\ref{singtrac}) of gauge-theory amplitudes.

\end{document}